# Excitation transfer from Second to First resonance line of Potassium observed in hot atomic vapor

**C. Andreeva[1], A. Krasteva[1], A. Markovski[2], S. Tsvetkov[1], S. Gateva[1], S. Gozzini[3], S. Cartaleva[1]**

**Abstract** We present experimental investigation on the fluorescence profiles observed by excitation of the hyperfine transitions of the second resonance line of potassium with a wavelength of 404.4 nm in dependence on the atomic density. This leads to both direct decay of the excited level population to the ground state (violet fluorescence), and to cascade decay via the first resonance lines (infrared fluorescence). It has been shown that the behavior of these two fluorescence profiles is different: increasing the atomic density, the violet fluorescence profile exhibits a well-pronounced self-absorption dip, while the infrared line does not show any narrow-width reduced absorption structure. Moreover, the profiles of the infrared line have a higher signal-to-noise ratio than that of the violet line. Our investigations show that beside atomic population, atomic polarization is also transferred by the cascade transitions. This is evidenced by registration of coherent magneto-optical resonances at the two fluorescence lines. The signal-to-noise ratio of these resonances registered at the first resonance line is significantly higher than at those obtained at the second resonance line. The proposed study makes it possible to examine cascade transitions in alkali atoms, particularly the preservation of atomic polarization, i.e. the coherence transfer by cascade transitions.

| | |
|---|---|
| Christina Andreeva | https://orcid.org/0000-0001-5549-2485 |
| Stoyan Tsvetkov | https://orcid.org/0000-0002-5216-3990 |
| Sanka Gateva | https://orcid.org/0000-0003-1270-5932 |
| Silvia Gozzini | https://orcid.org/0000-0002-5818-6456 |
| Stefka Cartaleva | https://orcid.org/0000-0002-1518-7790 |

[1] Institute of Electronics, Bulgarian Academy of Sciences, 72 Tzarigradsko Chaussee blvd., Sofia 1784, Bulgaria
[2] Technical University of Sofia, 8 Sv. Kliment Ohridski blvd., Sofia 1756, Bulgaria
[3] Istituto Nazionale di Ottica del CNR - SS Pisa, Via Moruzzi 1, Pisa, Italy

## 1 Introduction

One of the most interesting phenomena, based on the long living ground-state coherence, is the Coherent Population Trapping (CPT) (Alzetta et al. 1976; Arimondo 1996). The cancellation of resonant light absorption under CPT conditions is a manifestation of the so called effect of Electromagnetically Induced Transparency (EIT) (Harris 1997). The steep dispersion observed due to Zeeman coherence results in ultra-slow group velocity for a light pulse propagating in the atomic vapor (Budker et al. 1999) and also to various magneto-optical polarization effects. Such coherent interaction allows efficient control of the properties of the individual atoms by means of light field(s). One of the most important features of these coherent phenomena is the strong enhancement of the nonlinearities with the density of the resonant atoms. For example, the nonlinear Faraday rotation in hot atoms is more than two orders of magnitude larger than at room temperature (Novikova 2004), and such large rotation corresponds to a large index of refraction of the medium.

It is often assumed that atoms interact with an electro-magnetic field independently of each other, and have no influence on each other. However, such assumption is good only for optically thin media. The increase in atomic density can lead to collective phenomena such as, for example, local atomic field interactions and atomic collisions. The radiation trapping is one of these collective effects and is probably the most sensitive to the density of the atomic medium. Since the re-absorbed and spontaneously emitted photons are dephased and depolarized with respect to the coherent laser field creating the atomic polarization, the effect of radiation trapping can be described as an external incoherent pumping of the atomic transitions (Matsko et al. 2001, Novikova 2004). Using this model, the modifications of the CPT and the nonlinear Faraday resonance profiles due to radiation trapping have been calculated and it is shown that they lead to a decrease in the ground-state coherence.

Recently, potassium vapor contained in various optical cells has been successfully used for a very effective preparation of sub-natural-width coherent resonances on the first resonance $D_1$ and $D_2$ lines (Gozzini et al. 2009; Nasyrov et al. 2015; Lampis et al. 2016). The advantage of using K atoms is related to their relatively low ground-level hyperfine frequency difference (461.7 MHz), which is much less than the Doppler width of the optical transitions involved (~765 MHz at room temperature). In this case, the overlapping of the Doppler profiles of the transitions starting from both ground-state hyperfine levels can provide a good re-population of the ground-state level resonantly depleted by the light, thus enhancing the efficiency of the preparation of coherent population trapping or magneto-optical resonances (Gozzini et al. 2017). Such system can be considered as similar to a closed three-level system and appropriate for performing a proof-of-principle study devoted to simultaneous preparation of EIT resonances on the first and second resonance lines of potassium, by means of excitation of the second resonance line by a single frequency 404 nm light. This would make possible the systematic study of the cascade transfer of atomic coherence and polarization from the second to the first resonance line, which would allow a precise measurement of their dependence on potassium density. This is of great importance for the study of processes in optically thick alkali atom vapour, used in basic experiments related to hot atoms. The magneto-optical resonance preparation on the second resonance line and its registration on the $D_1$ and $D_2$ lines can be useful for various practical applications, due to the possibility to avoid the laser intensity noise in ultrahigh precision measurements, like the recently reported by Guidry et al. (2017).

In this communication, we present our experimental study of population transfer by registration of the fluorescence profiles observed by excitation of the hyperfine transitions of the second resonance line of potassium-39 ($4s^2S_{1/2} \rightarrow 5p^2P_{3/2}$) with a wavelength of 404.4 nm (further on referred as a violet line), in dependence on the atomic density. The energy levels involved are presented schematically in Fig. 1, where for simplicity only the relevant fine energy levels are shown.

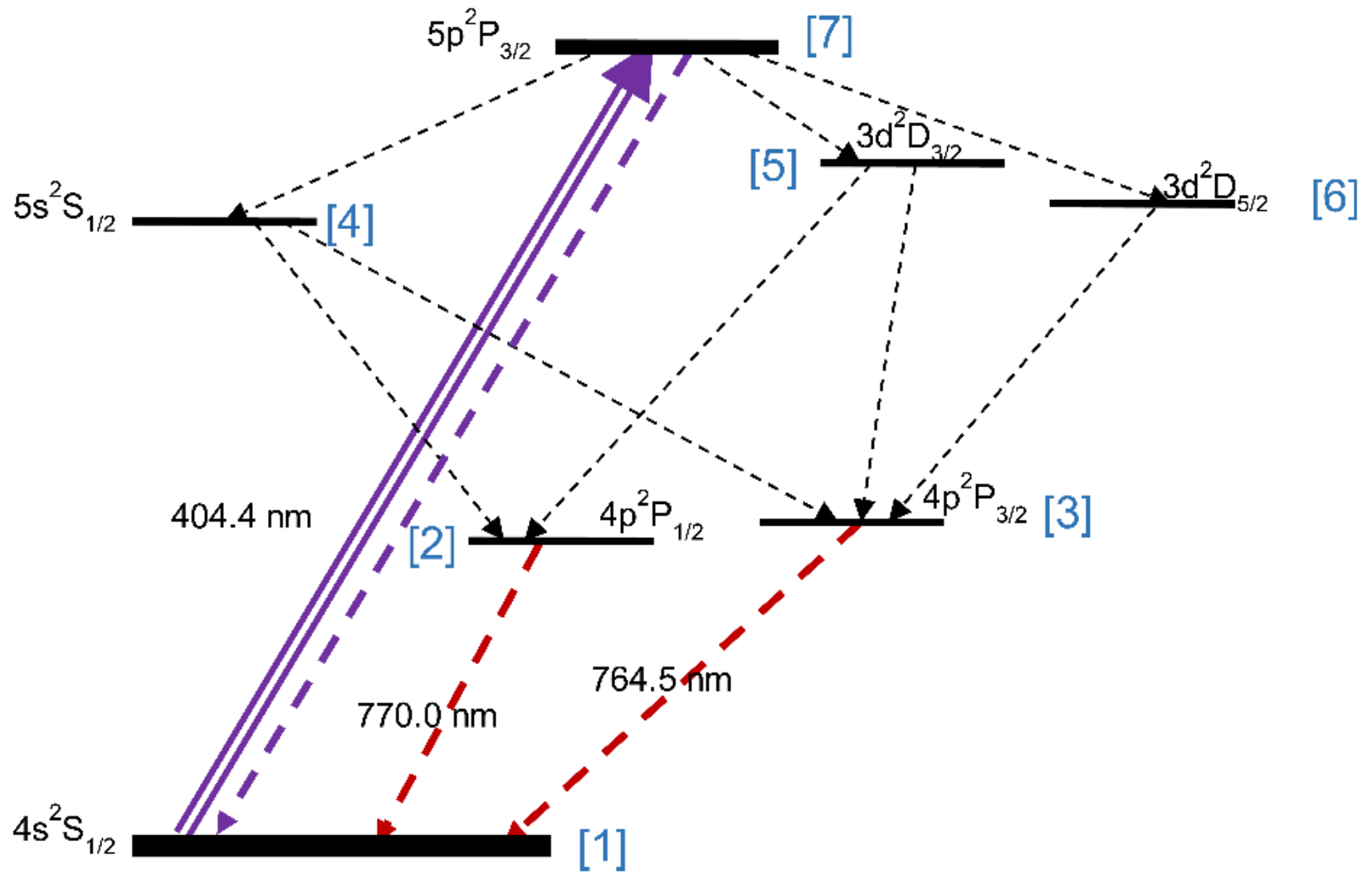

**Fig. 1** Scheme of the K fine energy levels involved

Such laser light excitation results in: (i) 404.4 nm fluorescence decay to the ground state and (ii) partial transfer of the $5p^2P_{3/2}$ atomic population to the excited $4p^2P_{1/2}$ and $4p^2P_{3/2}$ states of the first resonance line via a series of cascade transitions. The atomic population transfer is evidenced by registration of the fluorescence profile at the first resonance transition, referred as an infrared line. The fluorescence profiles show the influence of self-absorption in the hot media.

It should be noted that not only population, but also atomic polarization is transferred through the cascade transitions. To demonstrate this, the EIT resonances registered in the so called ground-state Hanle configuration at the violet and the infrared lines are compared. In the Hanle configuration only one traveling wave is needed and the absorption or fluorescence signal is a function of a static magnetic field applied along or orthogonally to the laser beam propagation. More specifically, at zero magnetic field the exciting linearly polarized light creates polarization (alignment) of the lower-level atomic population, and this alignment is destroyed in the presence of orthogonal magnetic field. Since long-lived ground-state magnetic sublevels are involved in the coupling scheme, their lifetimes determine the narrow linewidth of the obtained resonances.

## 2 Experimental setup

The scheme of the experimental setup is presented in Fig. 2. In the experiment, potassium vapor contained in an evacuated optical cell of diameter D = 2.5 cm and length L = 5 cm was excited by a tunable single-frequency, linearly polarized laser light, resonant with the violet transition. The fluorescence profiles were registered by a PhotoMultiplier System (PMS) separately for the violet and infrared lines, using two bandpass filters (centered at 405 nm and 770 nm respectively with 10 nm bandwidth). Thus, the violet and the infrared lines are well discriminated, but the infrared signal contains also 764.5 nm fluorescence. The laser frequency detuning was monitored by means of an auxiliary K reference cell (not shown in Fig. 2). An extended cavity diode laser (ECDL) was used, operating in single mode. The laser linewidth was of the order of 1 MHz. The laser power used in the experiment was 2 mW, and the laser beam cross-section was 0.02 $cm^2$. The optical cell was shielded against stray magnetic fields. When the K atoms are excited by linearly polarized 404.4 nm light, coherent superposition of ground state Zeeman sublevels is induced at $B = 0$, which was registered by measuring the 404.4 nm fluorescence as a function of the magnetic field. In this way, EIT resonances were observed in Hanle configuration, monitoring the fluorescence dependence on an orthogonal to the atomic beam magnetic field B, varied around B = 0. The field was provided by Helmholtz coils (not shown in Fig. 2) situated within the magnetic shield.

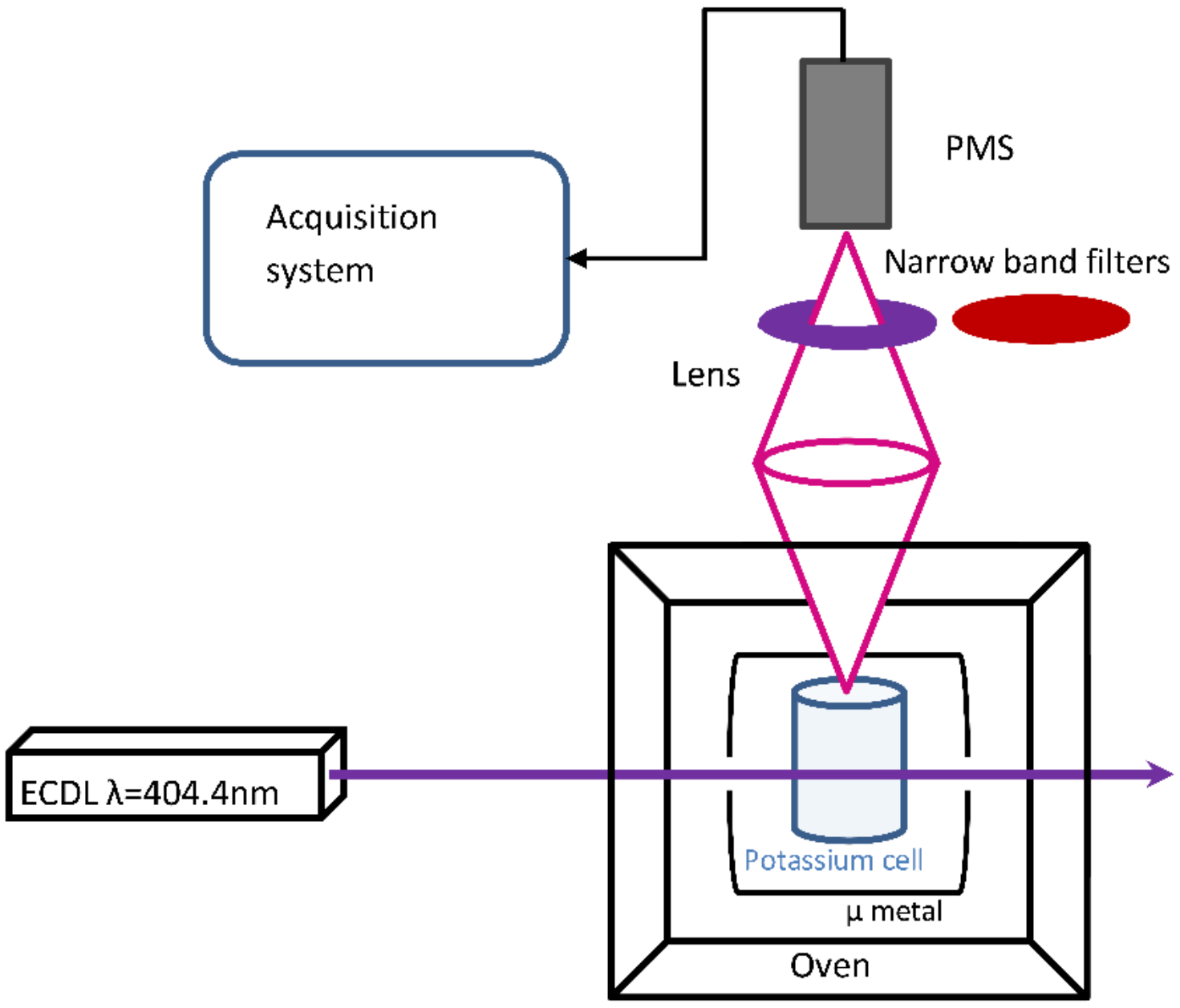

**Fig. 2** Scheme of the experimental setup

## 3 Experimental results

First, the Doppler broadened fluorescence profiles of the violet and infrared lines were measured for different atomic source temperatures in the interval from $T_{cell} = 117^oC$ to $T_{cell} = 172^oC$ (Fig. 3). With the enhancement of potassium source temperature, the violet fluorescence profile exhibits a well-expressed self-absorption dip, shown in Fig. 3a. This is different from the behavior of the infrared fluorescence line, which does not show any narrower-width reduced absorption structure, as seen from the results presented in Fig. 3b.

It is worth noting that the violet fluorescence profiles are superimposed on a background of 404 nm scattered laser light, whose level in this case is 25% of the maximal fluorescence measured at $T_{cell} = 117^oC$. Moreover, the scattered light background shows a significant increase with atomic density, when measured in the far wings of the respective Doppler profiles (Fig. 3a). The dependence of the level of the scattered 404 nm light on the potassium density shows that this scattering cannot be attributed only to the imperfect optical quality of the cell windows and walls. In experiments with hot (optically thick) alkali vapor where numerous acts of fluorescence light re-absorption and re-emission take place, the number of 404 nm photons propagating in different directions within the volume of the optical cell can be enhanced significantly.

In the case of the infrared fluorescence profile, no background of scattered laser light is registered and all five fluorescence profiles start from the zero-signal level (Fig. 3b), which is measured when the 404 nm laser beam is blocked. In addition, the experimental registration of the fluorescence profiles of the infrared line demonstrates a higher signal-to-noise ratio than that of the violet line, particularly for higher atomic source temperatures. The presented in Fig. 3a,b experimentally measured zero signals at closed 404 nm laser beam, show that the noise of the registration system is similar for both the violet and the infrared spectral line measurements.

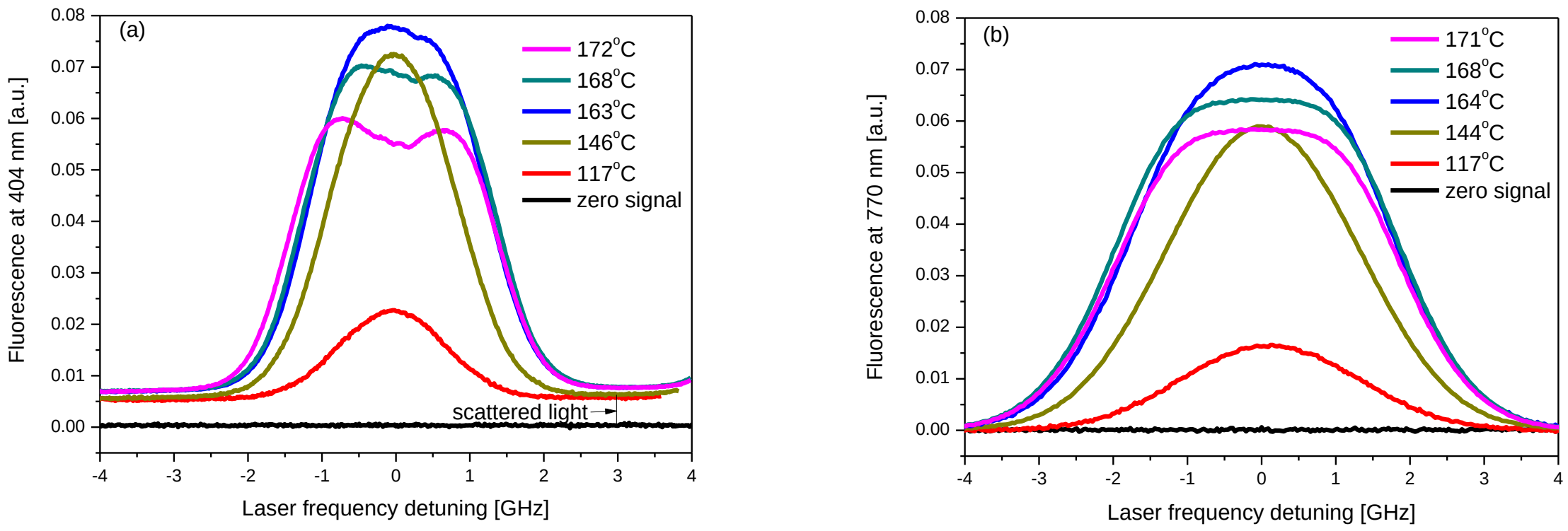

**Fig. 3** Fluorescence profiles of (a) the violet spectral line and (b) the infrared spectral line, at different optical cell temperatures.

Our experimental study has shown that the laser intensity noise can significantly compromise the measurements of the low-amplitude 404 nm fluorescence profiles for relatively high atomic source temperatures. Hence, the registration of the violet fluorescence suffers significantly from the noise due to the scattered laser light mixed with the fluorescence signal. But the situation is better when the proposed approach of infrared fluorescence registration is applied, since the exciting 404 nm laser light is completely cut off by the optical filter.

The amplitudes of the violet and the infrared lines follow a similar temperature dependence. More specifically, the amplitudes at resonance center increase, reaching a maximum at about 150°C and then decrease, as shown in Fig. 3. For a larger temperature interval, this behavior is shown in Fig. 4 and it is in agreement with the results reported by Matsko et al. (2001), where two different regimes for the incoherent pumping rate of the fluorescence photons are distinguished. More specifically, at low atomic densities this rate increases linearly with density due to the absorption and emission of photons within the cell volume. The other regime occurs for densities when photon reabsorption becomes significant inside the laser beam.

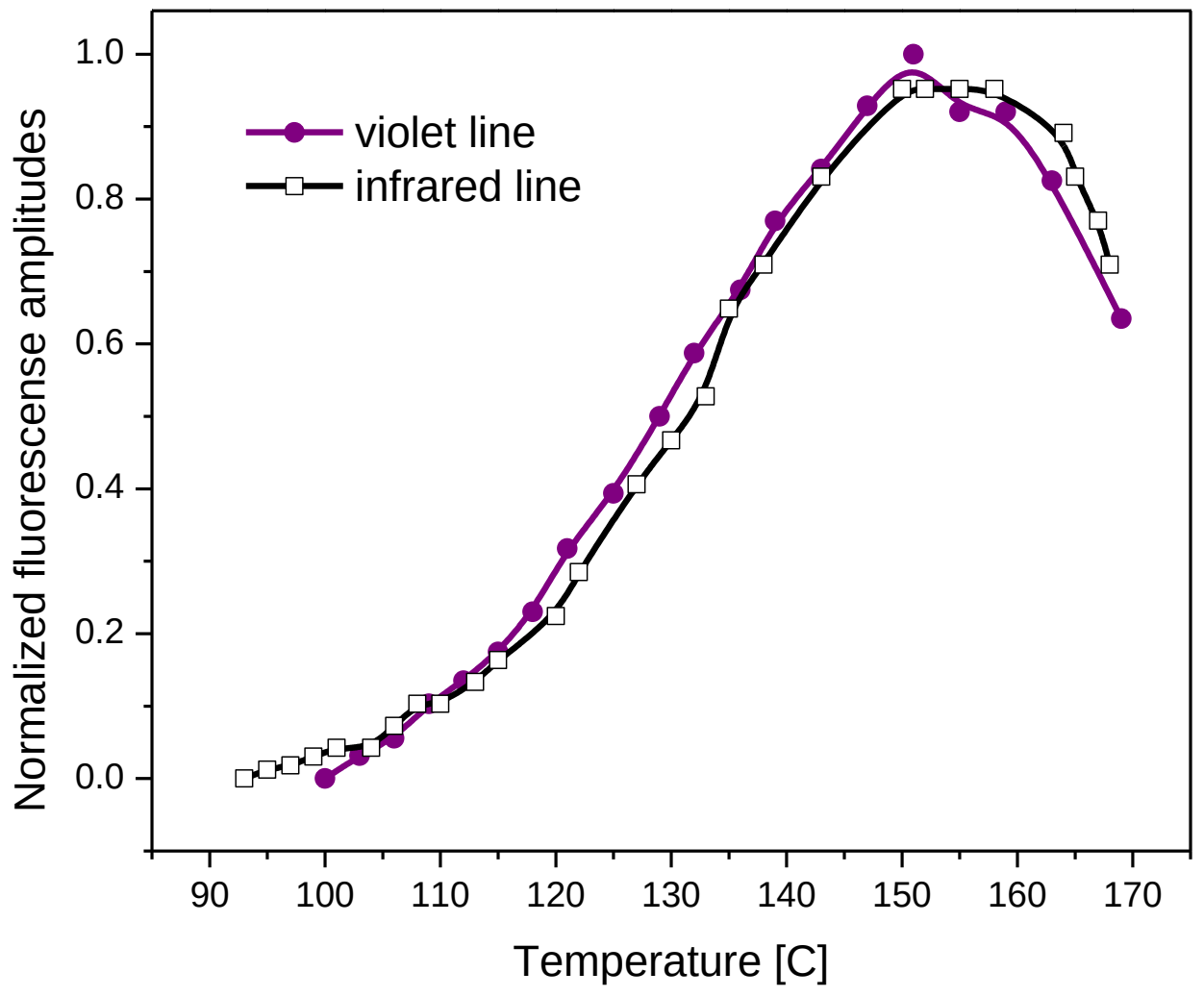

**Fig. 4** Dependence of the fluorescence amplitudes on the cell temperature

Having evidenced the different behavior of the fluorescence channels for hot media, it is important to investigate whether the transfer of atomic polarization is also dependent on the relaxation pathway. For this purpose, the same optical cell containing pure potassium vapor was excited by the linearly polarized laser light at 404.5 nm, and magneto-optical resonances in the fluorescence were observed in Hanle configuration, when an orthogonal to the laser beam magnetic field B was scanned around B = 0 (Gozzini et al. 2009, Gozzini et al. 2009a).

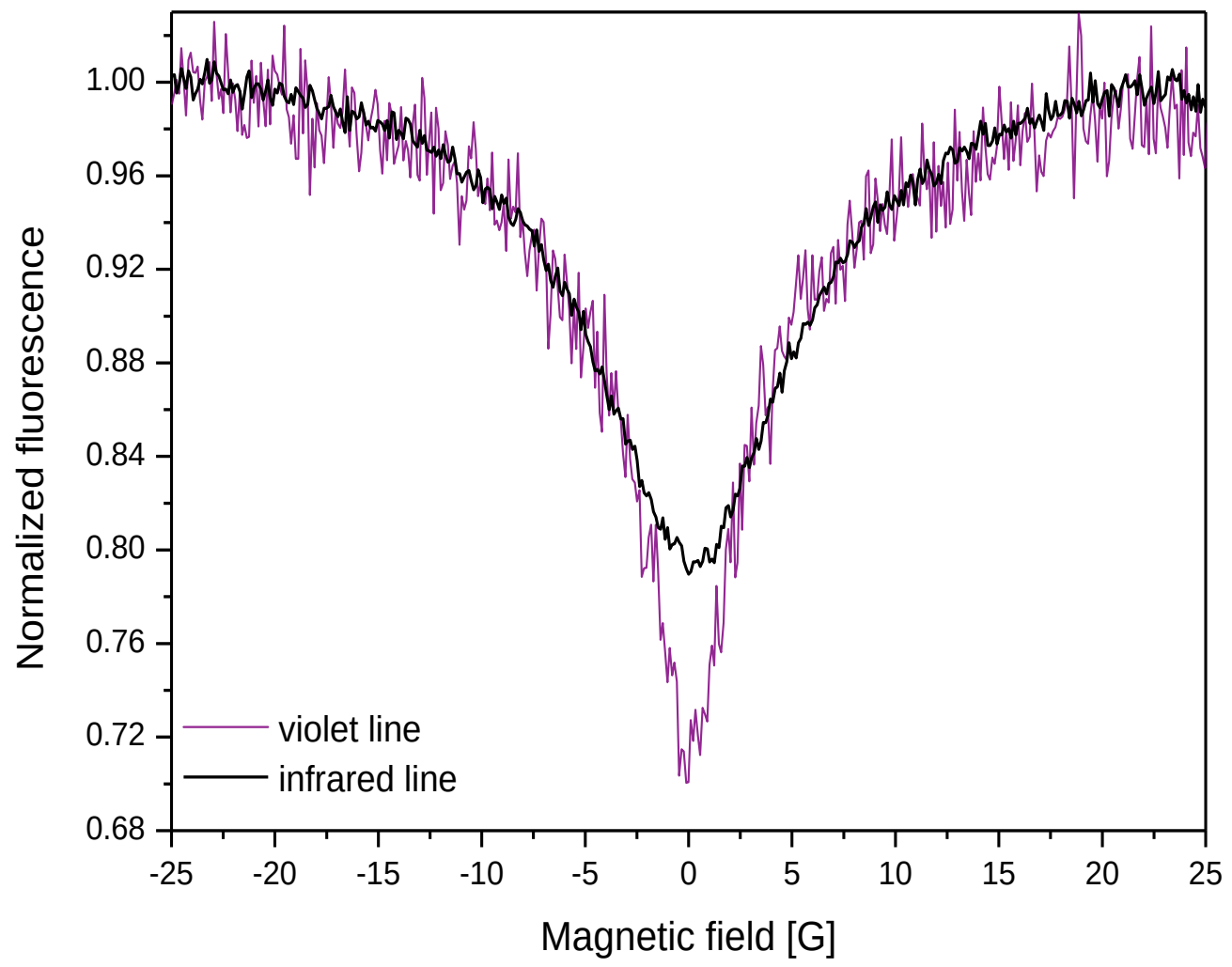

**Fig. 5** Magneto-optical resonances registered on the violet spectral line and the infrared spectral line fluorescence

The advantages in terms of signal-to-noise ratio of the infrared signal registration are more significant when magneto-optical resonances are studied, due to the fact that their amplitude is lower: typically (10-20)% of the amplitude of the Doppler profile. Fig. 5 presents the magneto-optical resonances at the violet and the infrared line for the case of $T_{cell} = 137^oC$. It can be seen that the noise characteristics of the two signals are different – the resonance measured on the violet fluorescence has a signal-to-noise ratio S/N = 7, while the one observed on the infrared fluorescence is with a higher S/N = 18.

Another important feature of these signals is the difference in their shape and width. As can be seen, the wings of violet and the infrared profiles coincide, but the 404 nm profile has a narrow structure in the center. As a result, the less noisy resonance at the infrared line appears broader (FWHM = 12 G) than that observed at the violet line (FWHM = 7 G).

This result will be studied further and it opens a new possibility for the investigation of the atomic polarization and coherence transfer by spontaneous transitions depending on reabsorption process and atomic density in hot atoms.

## 4 Phenomenological model

In order to analyze the observed differences, a simple model describing the atomic population transfer between relevant potassium levels was developed. The system considered includes the fine levels presented in Fig. 1, excited by linearly polarized laser light resonant with the fine transition $4^2S_{1/2} \rightarrow 5^2P_{3/2}$ ([1] → [7] according to the notation in Fig. 1) with a wavelength of 404.4 nm. The population transferred to level [7] then decays spontaneously both directly and via cascade transitions back to the ground state. The possible cascade transitions are marked by dashed lines in Fig. 1. The values of the corresponding branching ratios $b_{ij}$ for the transitions $i \rightarrow j$, as well as the fine levels lifetimes $\tau_i$ determining the natural linewidths $\gamma_i$ of the transitions have been taken after (Nandy et al. 2012). The loss of population of the ground state has been considered as in (Gozzini 2009a; Dancheva 2014) by introducing a finite ground-state linewidth of $\Gamma = 10$ MHz. Thus, the steady-state system of equations for the case of homogeneous (natural) linewidths of the levels is:

$$\frac{dn_1}{dt} = 0 = (n_7 - n_1)\sigma_{17} + n_7.b_{71}.\gamma_7 + n_2.b_{21}.\gamma_2 + n_3.b_{31}.\gamma_3(N - n_1)\Gamma \quad (1)$$
$$\frac{dn_7}{dt} = 0 = -(n_7 - n_1)\sigma_{17} - n_7.\gamma_7 - n_7.\Gamma$$
$$\frac{dn_4}{dt} = 0 = n_7.b_{74}.\gamma_7 - n_4.\gamma_4 - n_4.\Gamma$$
$$\frac{dn_5}{dt} = 0 = n_7.b_{75}.\gamma_7 - n_5.\gamma_5 - n_5.\Gamma$$
$$\frac{dn_6}{dt} = 0 = n_7.b_{76}.\gamma_7 - n_6.b_{63}.\gamma_6 - n_6.\Gamma$$
$$\frac{dn_2}{dt} = 0 = n_4.b_{42}.\gamma_4 + n_5.b_{52}.\gamma_5 - n_2.b_{21}.\gamma_2 - n_2.\Gamma$$
$$\frac{dn_3}{dt} = 0 = n_4.b_{43}.\gamma_4 + n_5.b_{53}.\gamma_5 + n_6.b_{63}.\gamma_6 - n_3.b_{31}.\gamma_3 - n_3.\Gamma$$

Here, $n_i$ is the steady-state population of level $[i]$ (Fig.1), and $\sigma_{17}$ is the excitation rate of the $4^2S_{1/2} \rightarrow 5^2P_{3/2}$ ([1] → [7]) transition, expressed as:

$$\sigma_{17} = \left(\frac{\gamma_7}{2} s\right)\left[\frac{(\gamma_7/2)^2}{(\nu - \nu_{17})^2 + (\gamma_7/2)^2}\right] \quad (2)$$

The value of $N$ is the atomic density, so initially $n_1 = N$, and $n_{i\neq 1} = 0$. The parameter $s$ is the so called saturation parameter, determined as $s = I/I_{sat}$, with $I_{sat}$ being the saturation intensity of the optical transition. In our calculations, we take $s = 1/2$. The solutions of this system of equations give the steady-state populations of all seven fine levels. In our case, we are interested in the fluorescence profiles of the second and the first resonance lines. The $D_1$ and the $D_2$ transitions have similar behavior in terms of lifetime, excitation rate and self-absorption threshold. Therefore, for simplicity, we consider only the 404.4 nm and the 770 nm transitions, which corresponds to plotting the populations of levels $5^2P_{3/2}$ (i.e. [7]) and $4^2P_{1/2}$ (i.e. [2]) vs the laser frequency detuning.

The next step is to introduce the inhomogeneously broadened Doppler profiles of the transitions. For this, the atomic population is expressed as:

$$N = N_0 exp\left(-\frac{(\nu-\nu_C^D)^2}{2\,\nu_D^2}\right) \tag{3}$$

where $\nu_C^D$ is the center of the Doppler transition and $\nu_D$ = 1.6 GHz is the Doppler width. Thus, the Doppler-broadened profiles for the fluorescence starting from levels [7] and [2] are obtained, for excitation of the violet transition.

With temperature increase, the atomic density in the cell increases. This leads to two effects. On one hand, there are more absorbing atoms, hence the fluorescence amplitude increases. On the other hand, the atomic layer between the interacting with the laser light atoms and the photodetector also becomes more dense and more re-absorbing the emitted incoherent resonant fluorescence light. According to Matsko et al. (2001), the threshold for observing the effect is determined by the relation:

$$\frac{3}{8\pi} N \lambda^2 D \frac{\gamma_i}{\omega_D} > 1 \tag{4}$$

In this expression, $\lambda$ and $\omega_D$ are the wavelength and the Doppler width (expressed as angular frequency) of the corresponding violet or infrared transition. The threshold corresponds to the probability of photon re-absorption becoming significant when the medium is optically thick on the length scale of the cell (in our geometry D = 2.5 cm).

From this relation it follows that the threshold for re-absorption for the infrared transition is much lower than that for the violet transition.

In our phenomenological model, we follow the approach of Matsko et al. (2001) and consider a two-level system coupled to a photon reservoir containing a thermal average photon number $\bar{n}_{th}$. At low temperature, the atomic density is low (optically thin medium) and $\bar{n}_{th} = 0$. Above threshold, though, the medium becomes optically thick and $\bar{n}_{th} > 0$. This thermal average photon number depends also on the laser frequency detuning, since at the wings of the Doppler profile the number of incoherent pumping photons is much smaller than at resonance. The two-level absorption is then introduced for both the violet and the infrared transitions, taking into account their different transition probabilities and saturation intensities. The results are presented in Fig. 6, for the case of the violet (a, b) and the infrared (c, d) fluorescence, respectively. Graphs (a) and (c) show the lines at 100°C, and graphs (b) and (d) correspond to a temperature of 170°C.

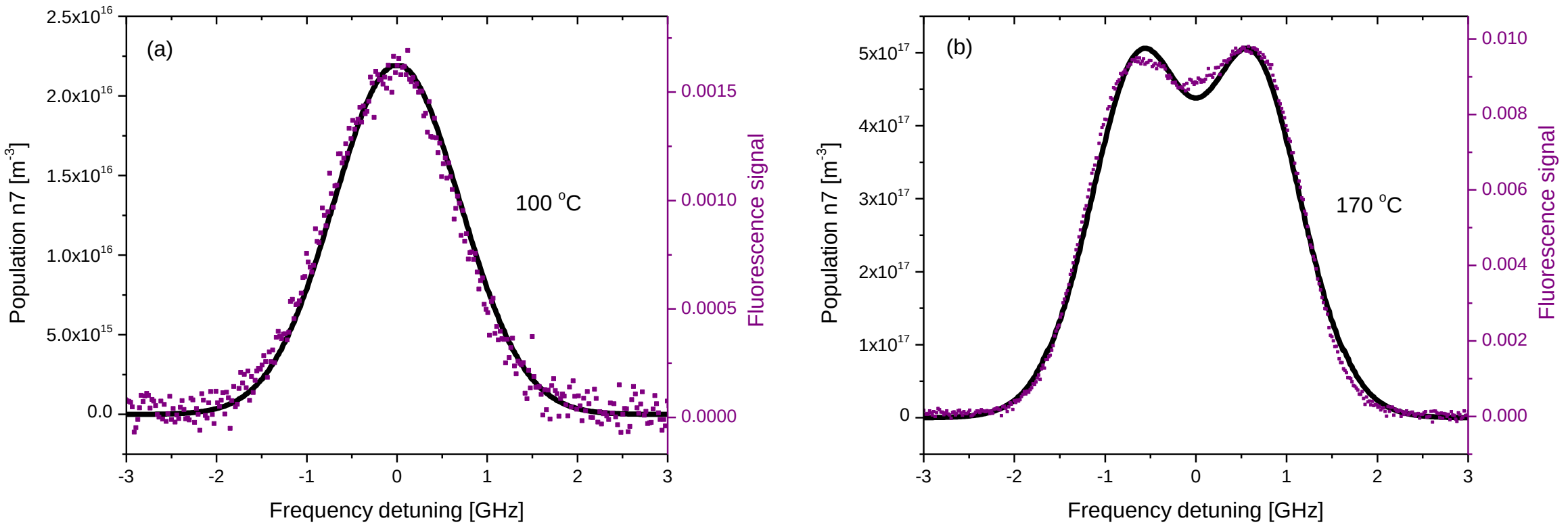

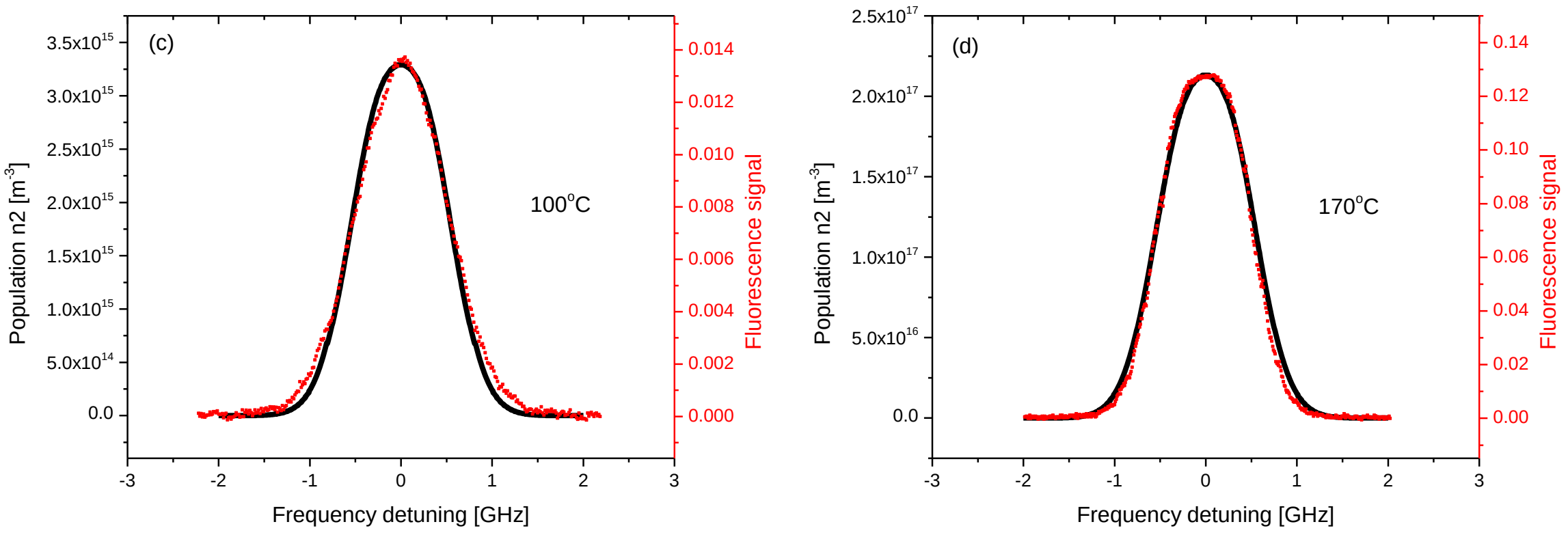

**Fig. 6** Fluorescence profiles of the violet (a, b) and infrared (c, d) fluorescence after re-absorption, for temperatures 100°C (a, c) and 170°C (b, d). Dots – experimental data, lines – numerical model

The slight asymmetry observed in the experimental dip in the violet fluorescence at high temperature is due to the hyperfine structure of the transition, which is not taken into account in the model.

It can be seen that the theoretical modeling is in good agreement with the experimental results (Fig. 6). More specifically, at lower atomic density, the thermal average photon number is low for the realization of a measurable re-absorption process and both the violet and the infrared lines show typical Doppler profiles. At higher density, the difference between the violet and infrared line behavior can be attributed to the significantly lower re-absorption threshold and higher absorption rate for the infrared line. Due to this, for the temperature interval considered, the re-absorption process at the 770 nm line is realized at a larger spectral interval around the resonance wavelength than it is in the case of the 404 nm line. In this way, the spectral region of re-absorption becomes comparable with the 770 nm line Doppler profile, and no self-absorption dip is formed.

## 5 Conclusions

In this paper we have presented our experimental investigation of the fluorescence profiles observed by excitation of the hyperfine transitions of the second resonance line of potassium ($4s^2S_{1/2} \rightarrow 5p^2P_{3/2}$) with a wavelength of 404.4 nm for different atomic densities. Such laser light excitation leads to both direct decay of the excited level population to the ground state (monitored through the 404.4 nm fluorescence), and to decay to the excited $4p^2P_{1/2}$ and $4p^2P_{3/2}$ states of the first resonance line, via cascade transitions (evidenced by registration of the first resonance line infrared fluorescence). It has been shown that the behavior of these two fluorescence profiles is different: increasing the atomic density, the violet fluorescence profile exhibits a well-pronounced self-absorption dip, while the infrared line does not show any narrow-width reduced absorption structure. Moreover, the fluorescence profiles of the infrared line have a higher signal-to-noise ratio than those of the violet line. Our experimental study has shown that this characteristic is also evidenced by the registration of EIT resonances in Hanle configuration. The signal-to-noise ratio of these coherent resonances registered at the infrared line is significantly higher than for those obtained at the violet line. This difference is attributed to the discrimination of the scattered laser light in the optical registration scheme. The proposed study makes it possible to examine cascade transitions in hot alkali atoms, particularly the preservation of atomic polarization, i.e. the coherence transfer by cascade transitions. It is also possible to study the light scattering processes in hot atomic vapors with the important practical outcome of avoiding laser light noise in the registration of weak coherent resonance signals.

One of the exciting potential applications of EIT and slow and stored light is for practical realization of quantum sensor technologies, quantum memory, and, ultimately, quantum computing. Warm atomic ensembles can be as practical as cold atoms for coherent manipulation of atomic spins using EIT. In addition, the warm-vapor-cell experiments have several attractive features, including a relative simplicity of the design and easy control over large atomic ensembles. Recent efforts to realize a high-efficiency memory using slow and stored light based on EIT in ensembles of warm atoms in vapor cells are reviewed by Novikova et al. (2012). We believe that the developed approach, described in the present paper based on the first and second resonance line in potassium, can be useful for the study and control of collective optical processes in warm atomic ensembles.

The quantum memory development strategy requires coherent optical memory modules that can efficiently store the coherently prepared quantum states of light and retrieve them on demand, without additional noise, i.e. long lifetime and low added noise memory. For this purpose, a ladder level scheme in warm rubidium atoms at room temperature was recently proposed by Finkelstein et al. (2018). A strong pulse induces the coherent absorption of the signal pulse in Rb, mapping the signal field onto a spatial field of quantum coherence between the lower and upper levels. A subsequent control pulse retrieves the signal via stimulated emission. In order to obtain a low noise in the registration, the scattered control laser light and spontaneous emission were filtered out spectrally. Thus, the further use of warm alkali atom ensembles will require elaborated systems for efficient laser and fluorescence noise suppression.

The strong reduction of optical cell longitudinal dimension (Sarkisyan et al. 2001) results in the observation of new phenomena when L approaches the wavelength $\lambda$ of the irradiating light. The experimental study of EIT resonances observed in optical cells with $L = 6\lambda$ has shown that their linewidth is significantly narrower than would be expected from the ground state dephasing rate due to atomic collisions with the cell windows (Cartaleva et al. 2012). Such linewidth is a sign that the thickness of $L = 6\lambda$ cell strongly reduces the probability of the fluorescence reabsorption by dense Cs atoms, due to the extremely short distance of the fluorescence photon within the cell.

The examples described above show the relevance of our investigations for the application of hot atoms as a medium for quantum control and sensing.

**Acknowledgements** This work was funded by the National Scientific Fund of Bulgaria, Grants DO08-19/2016, "New coherent and cooperative effects in hot alkali vapour" and DNTS/Russia 01/5/2017, "Nonlinear spectroscopy of spatially restricted alkali vapour: methodology and applications". ST and SG acknowledge the Bulgarian Academy of Sciences for the funding within Grant DFNP-17-76/2017. The authors are also grateful to Julia Fagan for proof-reading the manuscript.